\documentclass[entropy,article,accept,moreauthors,pdftex,10pt,a4paper]{mdpi} 
\firstpage{1} 
\articlenumber{x}
\doinum{10.3390/------}
\pubvolume{xx}
\pubyear{2017}
\copyrightyear{2017}
\externaleditor{Academic Editor: name}
\history{Received: date; Accepted: date; Published: date}

\graphicspath{{./figures/}}

\pdfoutput=1

\Title{Coupled DM Heating in SCDEW Cosmologies}

\Author{Silvio Bonometto $^{1,}$*$^{,\dagger}$ and
 Roberto Mainini $^{2,\dagger}$}

\AuthorNames{Silvio Bonometto and Roberto Mainini}

\address{%
 $^{1}$ \quad {Istituto Nazionale di Astrofisica} (INAF), Trieste Astronomical Observatory, Via Tiepolo 11,
 34143 Trieste, Italy; bonometto@oats.inaf.it\\ 
 $^{2}$ \quad  {Dipartimento di Fisica G.Occhialini, Universit\'a Milano-Bicocca}, 
  Piazza della Scienza 3,
 20116 Milano, Italy; mainini@mib.infn.it}

\corres{Correspondence: bonometto@oats.inaf.it}

\firstnote{These authors contributed equally to this work.}

\abstract{Strongly-Coupled Dark Energy plus Warm dark matter (SCDEW)
 cosmologies admit the stationary presence of $\sim$1$\%$ of
 coupled-{DM} and DE, since inflationary reheating. Coupled-DM
 fluctuations therefore grow up to non-linearity even in the early
 radiative expansion. Such early non-linear stages are modelized
 here through the evolution of a top-hat density enhancement,
 reaching an early virial balance when the coupled-DM density
 contrast is just 25--26, and the DM density enhancement is $ \sim$10$\,
 \%$ of the total density. During the time needed to settle in virial
 equilibrium, the virial balance conditions, however, continue to
 modify, so that ``virialized'' lumps undergo a complete evaporation.
 Here, we outline that DM particles processed by overdensities
 preserve a fraction of their virial momentum. Although fully
 non-relativistic, the resulting velocities (moderately) affect the
 fluctuation dynamics over greater scales, entering the horizon later
 on.}

\keyword{cosmology; dark matter; dark energy; non-linear fluctuation dynamics}

\begin{document}


\section{Introduction}

Almost two decades have elapsed since Hubble diagrams of Type Ia supernovae ({SNIa}) 
\cite{riess,perlmutter}
led to the cosmic acceleration discovery. {Lambda cold dark matter (LCDM)} 
models, formerly treated as
a counterexample, were then revitalized, as providing an excellent
data fit with a minimal extra parameter budget. Since then, cosmologist
lived a sort double life: From one side, more and more data were found
to fit LCDM, first of all the observed gap between total and matter
density parameters, \mbox{$\Omega_0 (\simeq 1)$} and $\Omega_{0m} (\simeq 0.3)$, that
WMAP  {({https://lambda.gsfc.nasa.gov/product/map/dr5})} 
and
Planck {(https://www.cosmos.esa.int/web/planck)} data made
sure. From the other side, they cannot ignore the extreme fine-tuning
and the coincidence conundrums that LCDM implies. The component or phenomenon
accounting for the density parameter $\Omega_{0d} = \Omega_{0} -
\Omega_{0m} (\simeq 0.7)$ was however dubbed Dark Energy (DE).

In the first decade of the new millennium, therefore, quite a few
ideas were suggested or revitalized, aiming to gain an insight into
the true DE nature. However, any possible option requires the
introduction of extra parameters, with respect to LCDM, so that a
significantly better data fit should balance such model ``degradation
''.
On the contrary, even the most successful options succeeded even
in the LCDM data fit, at most.

Within this context, the natural option was to suggest and plan new
experiments. In particular, it seems essential to find an independent
confirmation of cosmic acceleration, whose evidence still lies on SNIa data
alone. In the second decade, deep sky experiments, such as the Dark Energy Survey
({DES}) {(https://www.darkenergysurvey.org)} and
Euclid {(https://www.euclid-ec.org)}, were then planned. It
became soon evident, however, that such experiments are able to
discriminate LCDM only vs.~fairly extreme options, not vs.~close
models {(see, e.g., \cite{kleidis2015,kleidis2015a,kleidis2015b,Benisty2017,Benisty2017a,Haba2017})}.

Distinguishing General Relativity (GR) from more sophisticated
gravitational approaches or confirming the (non-)significance of
possible hidden dimensions will however be a basic success. Although
the expected confirmation of GR and the space dimension number will leave us
with the same conflict on DE nature from which we started.

Within this context, in this paper, we consider the Strongly-Coupled Dark Energy plus Warm dark matter (SCDEW) cosmology.
Running experiments could hardly help us to discriminate it vs.~LCDM.
However, SCDEW substantially eases LCDM's conundrums, first of all; its
main success however concerns scales below the average galactic
scale, where several data still disagree with LCDM or are fit just at
the expense of 
 {\it ad hoc} baryonic physics.

SCDEW cosmologies were already discussed in quite a few previous
papers
\cite{bonometto2012,bonometto2014,bonometto2015,maccio2015,bonometto2017a,bonometto2017b}
 {showing that: (i) In these models, the inflationary period
  ended in an era of Conformally-Invariant (C.I.) expansion,
  when Dark Matter ({DM}),
  DE and radiation had constant early density
  parameters; this era approaches an end when fields and
  particles, namely DM particles, acquire a Higgs' mass. While
  some parameter values need to be suitably selected to meet the
  observations, no fine-tuning or coincidence problem remains,
  apart from the similarity between the present baryon and DM
  abundances, a problem shared even by ancient ``standard'' CDM
  models. See Figure \ref{figure 1}, below, for a visual illustration of
  these points, which are the basic findings allowing for SCDEW
  models. (ii) Then, below the Higgs' scale, SCDEW models allow
  for DM components with masses $m_w \sim 100\, $eV. Ordinary
  LWDM models, with such DM masses, yield no structure below
  $\sim$10$^{12} h^{-2} M_\odot$. On the contrary, in SCDEW
  models, the low mass limit for structure formation shifts below
  the stellar mass range. In turn, it has been known for a long time
  that a low DM mass eases LCDM problems, like dwarf galaxy
  profiles, MW and M31	satellite scarcity and concentration
  distribution. For an exhaustive bibliography, see, e.g., 
  \cite{maccio2015}. A recent related analysis, based on NIHAO (Numerical Investigation of a Hundred Astrophysical Objects) 
  hydro simulations, can be found in
  \cite{maccio2017,Frings2017,Santos2017} and the papers cited
  therein. (iii) In general, the main discrepancies between
  SCDEW and LCDM predictions concern fairly small scales,
  typically below $\sim$10$^{10} h^{-2}M_\odot$, where SCDEW
  predicts that existing systems formed earlier than in LCDM.

 In SCDEW cosmologies,
DE is a (self-interacting) scalar field $\Phi$. It is then worth
specifying: (a) why SCDEW is intrinsically different from widely-studied {quintessential} models (see,
e.g., \cite{amendola2010,bamba} and the references therein); (b) which
aspects of SCDEW cosmologies will be deepened here.

As far as (a) is concerned, let us outline that the scalar
field $\Phi$, in SCDEW cosmologies, has been a significant cosmic
component since ``ever''
; after inflation, in fact, they predict a long
era of Conformally-Invariant ({CI}) 
expansion, when $\Phi$ accounts for
a constant share of the background cosmic budget, ranging between
permils and percents. This is so, while $\Phi$ is purely kinetic, thanks
to energy exchanges with a Dark Matter (DM) spinor field $\psi$,
accounting exactly for the double of $\Phi$ density. The rationale of
this picture is further discussed in the next section.

(b) Then concerns the growth of such coupled-DM density
fluctuations, during the epoch of CI 
 expansion. In a previous work, we
followed its non-linear stages by the treating of a top-hat density
enhancement and showed that, while other component fluctuations are
dissipated or enter a sonic wave regime, the DM component undergoes a
peculiar multi-step process: (i) when entering the horizon, it undergoes
a rapid growth, eventually slowed down by the exit from the
relativistic regime, in spite of the fact that the DM contribution to the
cosmic budget is, at most, $\sim$1\%; (ii) positive fluctuations,
with an amplitude not too far from the average gradually reach a
non-linear regime, which we model through the top-hat analysis; (iii)
after growing up to a suitable (moderate) density contrast, however,
the top-hat reaches the conditions for virial equilibrium; then, the
most unexpected stage follows, as: (iv) virial equilibrium is not a
stationary condition, and virialized top-hats gradually dissolve.

This is so because binding effects weaken as the effective mass of
coupled-DM particles degrades. This, however, leaves us with a problem,
as a non-negligible fraction of the momentum acquired in virialized
lumps continues to have particles evaporating from it. In this work, we
therefore debate the effects of this energy input, occurring on all
scales, in a (semi)quantitative way. A priori, one could envisage a
sort of danger: that smaller scale fluctuation degradation should input a
non-negligible energy amount, so affecting the treatment of greater
scale top-hat fluctuations, entering the horizon later on.

\begin{figure}[H]
\centering
\begin{tabular}{c  c}
\includegraphics[width=6cm]{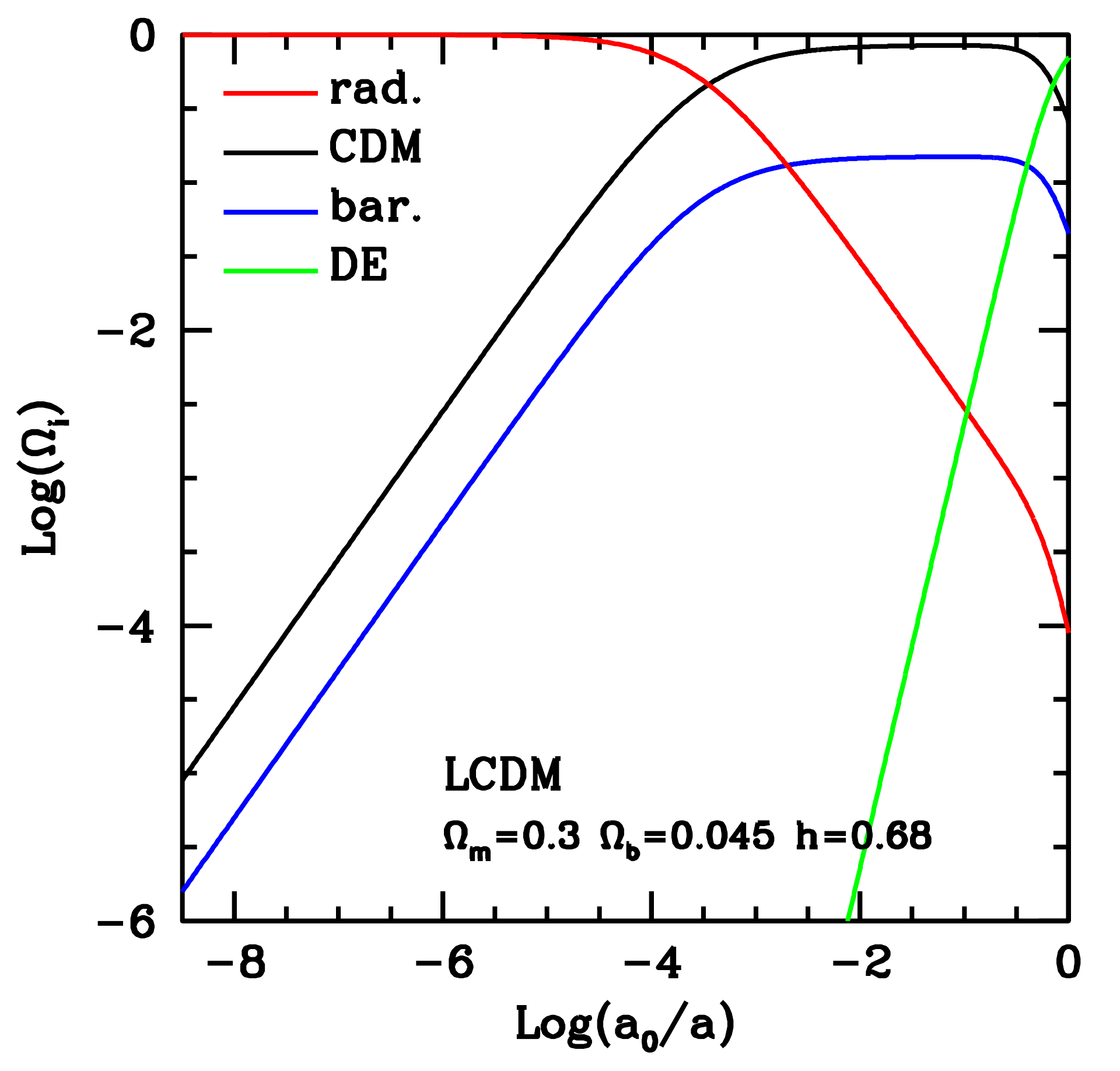}&
\includegraphics[width=6cm]{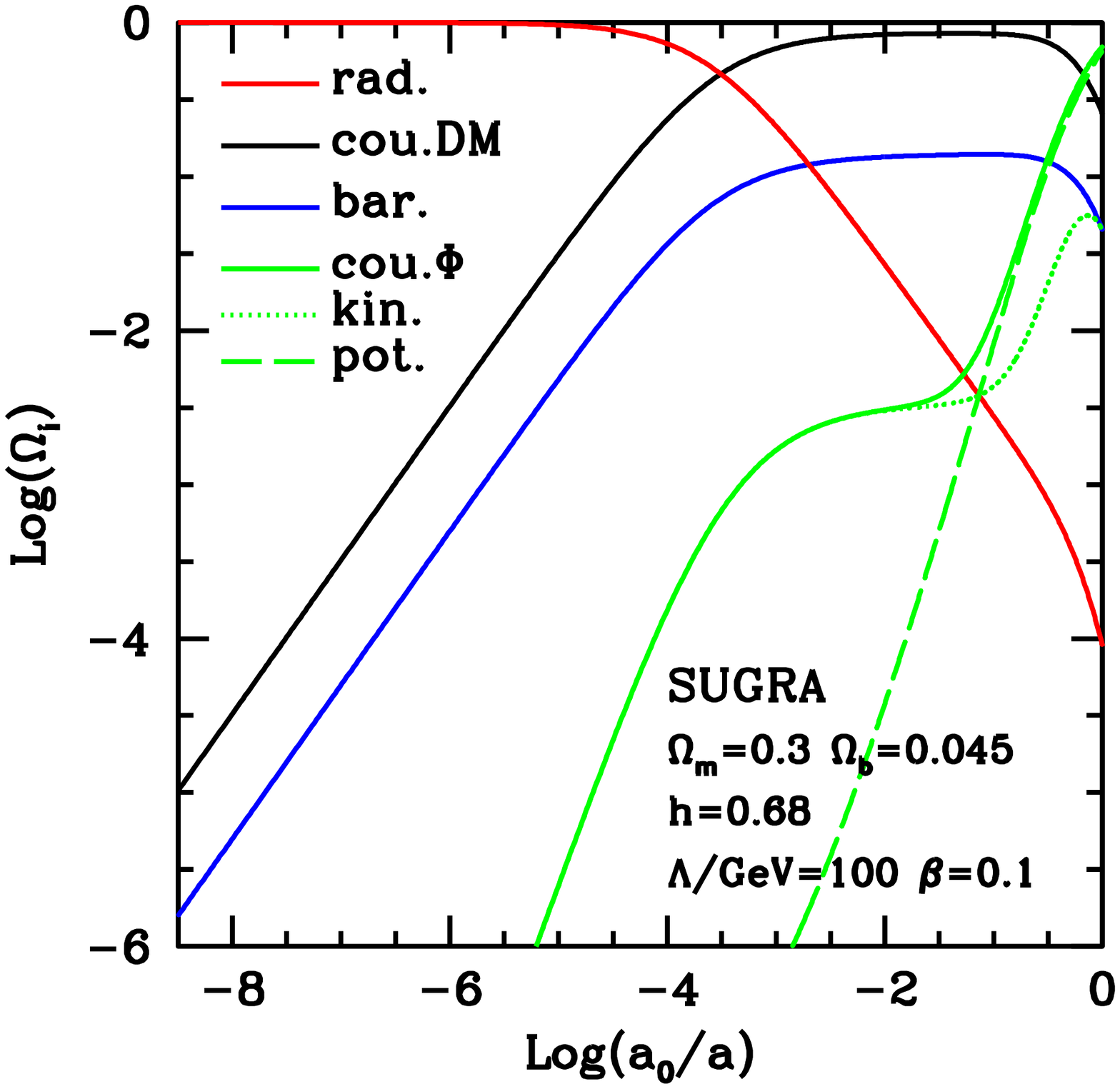}\\
({\bf a}) & ({\bf b})
\end{tabular}
\begin{tabular}{c}
\includegraphics[width=6cm]{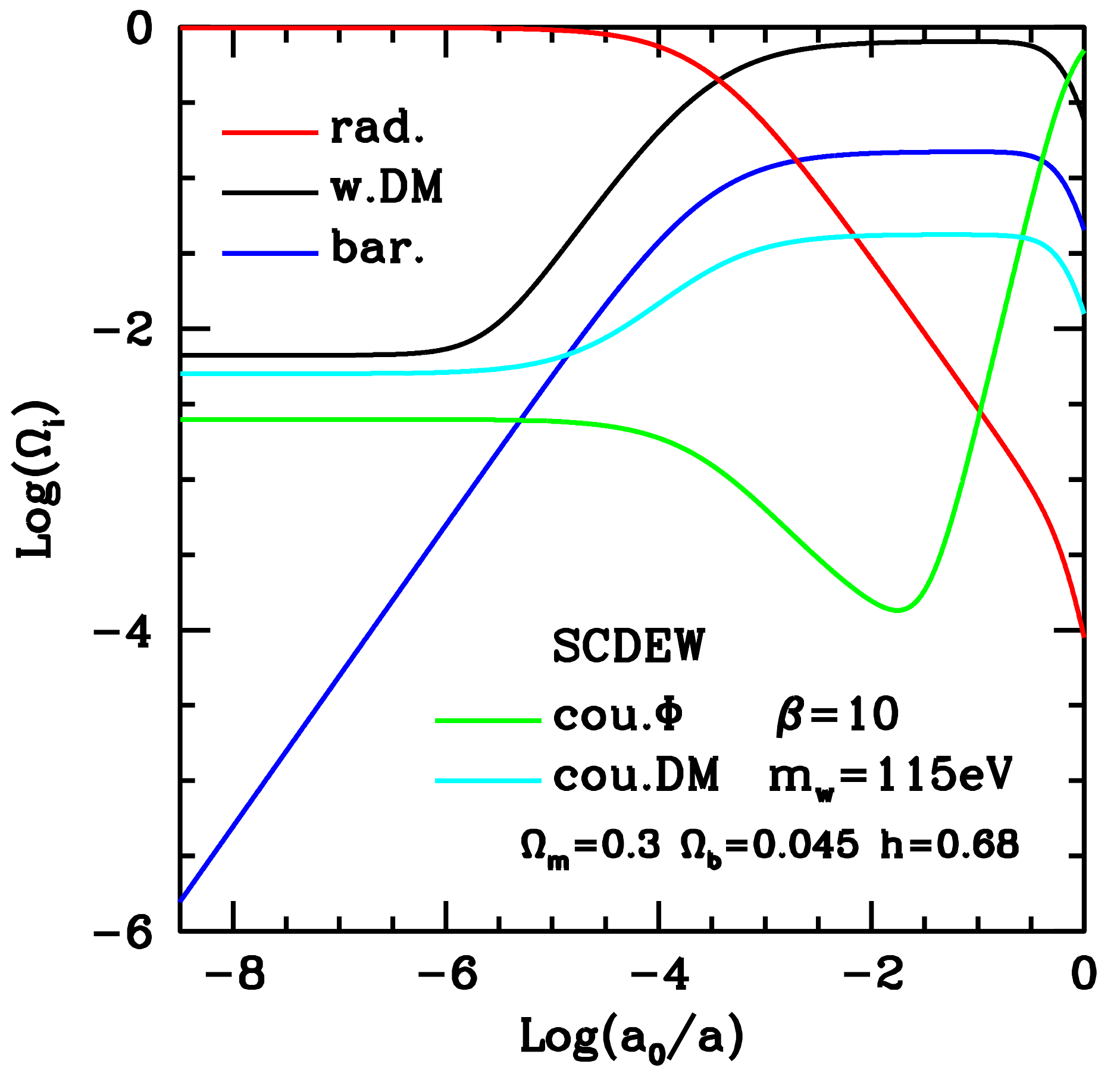}\\
({\bf c})  \\
\end{tabular}
\caption{Scale dependence of the density parameters $\Omega_i$ (the
 label refers to the different cosmic components) in: {(\bf a)} LCDM 
, {(\bf b)} coupled
 quintessence (with a {SUGRA}
 potential), {(\bf c)} Strongly-Coupled Dark Energy plus Warm dark matter (SCDEW) models. Model
 parameters in the frames. In the SUGRA plot, we also show the kinetic
 and potential components of DE density, so making clear where the
 potential-kinetic transition occurs. In the SCDEW plot, such
 transition takes place around the redshift where the green curve
 ($\Phi$ field energy) is minimum. As is known, the DE component
 rapidly vanishes when $1+z=a_0/a$ increases, both in LCDM and in the
 coupled SUGRA model, although in the latter case, its fall is delayed
 to $z > z_{eq}$ (matter-radiation equality). On the contrary, in
 SCDEW cosmologies, only baryons suffer a similar decrease: the plot
 clearly shows how, at high $z$, radiation, $\Phi$ and DM settle on
 parallel curves, in the late Conformally-Invariant (CI) expansion.} 
\label{figure 1}
\end{figure} 

Fluctuation dissipation occurs at different stages in the cosmic
evolution. When due to free streaming, it generally causes no energy
input. Sonic wave dissipation, at the epoch of matter-radiation
decoupling, instead, yields an energy transfer from cosmic to micro
scales. The actual impact of such energy input on cosmic evolution,
however, is nil.

As we shall see, in our case, the situation is different. The energy
input surely causes no change on the nature of cosmic components, but
the treatment of fluctuation evolution an scales entering the horizon
soon later could be (moderately) affected. As a matter of fact,
dealing with a spherical density enhancement, keeping to the
assumptions that materials start with no ``thermal'' energy, becomes a
simplifying approximation. It however allows us to estimate the size
of the {heating up} effect, for materials processed in low scale
fluctuations. The discussion on this effect is the main original
contribution of this paper.

Owing to the nature of the matter treated, however, we thought that a
self-consistent discussion requires us to resume the results of previous
papers. A further contribute of this paper is therefore an outline of
the physical context, which we shall present in an essential fashion.

The plan of the paper is therefore as follows: In Section \ref{sec2}, we show
the evolution of background densities in SCDEW cosmologies (Section \ref{bd}) 
and
debate linear fluctuation evolution (Section \ref{lf}). In Sections \ref{sec 3} and \ref{sec4}, we work
out the equations enabling us to follow the evolution of a top-hat
density enhancement. In Section \ref{sec5}, we discuss its virialization. The
equations worked out are numerically solved in Section \ref{sec6}. Sections \ref{sec7}
and \ref{sec8} then concern post-virialization events. Finally, in Sections \ref{sec9}
and \ref{sec10}, we discuss the results and draw our conclusions.

\section{The Peculiar DE Evolution in SCDEW Cosmologies: Background Densities }
 \label{sec2}

One of the motivations of cosmologies with DE being a scalar field
$\Phi$ was to allow DE to preserve a non-negligible density also at
$z >\sim 0.33$, where instead, $\Lambda$ begins to be be subdominant.
Partial success was achieved when DM was coupled with the field
$\Phi$, which so receives a continuous energy input and therefore
keeps a density that is an appreciable fraction of DM. Within
(almost) any such approach, $\Phi$ becomes purely kinetic when $z$ is
large enough; when this occurs, the energy feed from DM becomes
insufficient, and DE density rapidly falls down with increasing $z$.

In the first two plots of Figure \ref{figure 1}, we show typical behaviors of DE
and other density parameters in the LCDM and coupled quintessence
cases. In the third plot, densities are then shown for an SCDEW
cosmology. One immediately notices that, in the third plot, all
cosmic components (but baryons, alas!) keep significant densities in
all epochs, arising from parallel curves characterizing the early
CI expansion, lasting since ``ever''. The price to pay is the
simultaneous presence of two DM components, which, however, exhibit close
densities all through the C.I. expansion and, as better discussed
below, have equal Higgs' masses. Models with two DM components were
widely considered in the literature in an attempt to fit suitable
datasets. At variance from SCDEW, in these cases, a serious conceptual
problem arises, as one has to assume an (almost) coincidence of the
present density parameters for two components of different origin.
Altogether, the parameters to be added to LCDM are: (i) the value of
the mass of DM particles ($m_w$); (ii) the high-$z$ coupling intensity
($\beta$).

Also in SCDEW cosmologies, above a suitable redshift, the field $\Phi$
becomes purely kinetic. Instead of making use of a specific self-interaction potential, we simply input a transition redshift, where
the DE state parameter $w(a)$ shifts from $+1$ to $-1$ (see below). As a
matter of fact, $\Phi$ self-interaction parameters will be hard to
constrain by any foreseeable future experiment, while a detection of
$w(a)$ could be observationally easier. The rest of this section is
devoted to motivate these features.

\subsection{Background Dynamics}
\label{bd}

In this subsection, we shall resume the background analysis of an SCDEW
model. For linear inhomogeneities, we shall provide less details, which
can be however found in the previous papers on this subject
\cite{bonometto2014,bonometto2015,bonometto2017a}. Let us then use the
background metric:
\begin{equation}
ds^2 = a^2(\tau) (d\tau^2 - d\lambda^2)
\label{metric}
\end{equation}
$\tau$ being the conformal time, $d\lambda$ the line element, while
$a(\tau)$ is the scale factor. The state equation of a purely kinetic
scalar field $\Phi$, whose {free} Lagrangian reads:
\begin{equation}
{\cal L}_f \sim \partial^\mu \Phi \partial_\mu \Phi ~,
\label{freeLagrangian}
\end{equation}
is then $w=p_k/\rho_k \equiv 1$ ($p_k,~\rho_k:$ pressure, energy
density). Accordingly, $\rho_{k} = \dot \Phi^2/2a^2$ should dilute
$\propto a^{-6}$. It is also known that non-relativistic DM density
$\rho_c \propto a^{-3}$, its state parameter being $w=0$. It may then
appear intuitive that a suitable energy flow from DM to $\Phi$ could
speed up DM dilution while $\rho_k$ dilution slows down, so that both
dilute $\propto a^{-4}$.

As a matter of fact, in \cite{bonometto2012}, it was shown that these
expectations are consistent with a DM-$\Phi$ coupling ruled by the
equations:
\begin{equation}
T^{(d)~\mu}_{~~~\, ~~\nu;\mu} = +C T^{(c)} \Phi_{,\nu}~,
~~~~~~~~~~
T^{(c)~\mu}_{~~~~\nu;\mu} =- C T^{(c)} \Phi_{,\nu}~,
\label{relativisticeq}
\end{equation}
an option introduced since the early papers on DE (see
\cite{amendola2000} and the references therein) and yielding, e.g., the
model illustrated in the second panel of Figure \ref{figure 1}. In
Equation~(\ref{relativisticeq}), $T^{(\Phi,c)}_{\mu\nu}$ represents the
stress-energy tensors for the $\Phi$-field, DM, whose traces are
$T^{(\Phi,c)}$. The factor:
\begin{equation}
C = b/m_p= (16 \pi/3)^{1/2} \beta/m_p~
\label{couplingconstant}
\end{equation}
($m_p$: the Planck mass) causes a DM-DE coupling, therefore
parametrized by $b$ or $\beta$. When the Equations~(\ref{relativisticeq})
hold, in a radiation dominated epoch, DM and $\Phi$ densities
necessarily fall on an attractor, characterized by (constant) state
parameters:
\begin{equation}
 \Omega_\Phi = {1 \over 4 \beta^2}~,~~ \Omega_c = {1 \over 2 \beta^2}
 \label{earlyomegas}
\end{equation}
so that the requirement $\Omega_\Phi+\Omega_c \ll 1$ implies that
$\beta \gg \sqrt{3}/2$. Values of $\beta <\sim 2.5 $ are however
excluded by limits on {dark radiation} during {BBN} (Big Bang Nucleosynthesis)
 or when {CMB} (Cosmic Microwave Background)
spectra form.

In the frame where the metric is (\ref{metric}),
the Equation~(\ref{relativisticeq}) also read:
\begin{equation}
 \dot \Phi_1 + \tilde w{\dot a \over a} \Phi_1 = {1+w \over 2} C a^2 \rho_c
 ~,~~~~~~~ \dot \rho_c + 3 {\dot a \over a}\rho_c = -C \rho_c \Phi_1~,
\label{nonrelativisticeq1}
\end{equation}
with $\Phi_1 \equiv d \Phi/d \tau$ and $2 \tilde w = 1+3w - d \ln(1+w)
 / d\ln a$. By solving these equations, one obtains $\Phi_1$
evolution directly from $w(a)$ dependence. Notice that the system is
just second order, and $w(a)$ information always admits an arbitrary
additional constant~on~$\Phi$.

Equations~(\ref{relativisticeq}) or (\ref{nonrelativisticeq1}) are
obtainable, in a Lagrangian formulation, if DM is a spinor field
$\psi$ with a negligible kinetic term, while its interaction with
$\Phi$ arises from a generalized Yukawa interaction:
\begin{equation}
{\cal L}_m = - \mu f(\Phi/m) \bar \psi \psi
\label{interaction}
\end{equation}
provided that
\begin{equation}
f = \exp(-\Phi/m)~.
\label{fff1}
\end{equation}
Here, two mass scales, $m = m_p/b$ and $\mu = g\, m_p$, need to be
introduced for dimensional reasons.

 The constant $b$ coincides
with the factor $b$ gauging the DM-$\Phi$ interaction strength in
Equation~(\ref{couplingconstant}), so that $C=1/m$; on the contrary, $g$
keeps undetermined, 
as well as 
a $\Phi$ additive constant. {Accordingly, at any scale above the
 Higgs' scale, the $\Phi$ field Lagrangian shall amount to two terms:
 the kinetic term (\ref{freeLagrangian}) and the $\Phi$-$\psi$
 mass-interaction term (\ref{interaction}).}

As the particle number density operator of the spinor field $n \propto
\bar \psi \psi$, from the Lagrangian density (\ref{interaction}), we
see that the DM energy density reads:
\begin{equation}
 \rho_c = \mu f(C\Phi) \bar \psi \psi
 \label{rhoc}
\end{equation}
(formally, $= -{\cal L}_m$). It is then worth focusing on the term:
 \begin{equation} 
 {\delta {\cal L}_m \over \delta \Phi} \equiv [{\cal L}_m]'_\Phi = -
 \mu {f'}_\Phi (C\Phi) \bar \psi \psi = - {{f'}_\Phi (C\Phi) \over
 f(C\Phi)} \rho_c = C\rho_c
\label{eulerolagrange}
\end{equation}
of the Euler--Lagrange equation, which, multiplied by a suitable
factor, stands at the r.h.s. of the first
Equation~(\ref{nonrelativisticeq1}).

Let us now add that the equations~(\ref{nonrelativisticeq1}) can be soon
integrated, obtaining that:
\begin{equation}
 \Phi_1 = C/\tau~,~~~ \rho_c \propto a^{-4}
 \label{integre}
\end{equation}

\noindent
as $\Phi = C\ln(\tau)$, so that, in Equation~(\ref{rhoc}), $ f \propto
\tau^{-1} \propto a^{-1}$ during the radiative expansion, apart from an
additive constant. Of course, also:
\begin{equation}
 \rho_k = {\Phi_1^2 /( 2 a^2)}
\end{equation}
then dilutes as $a^{-4}$. This is why, in the radiative expansion,
$\Omega_\Phi$ and $\Omega_c$ keep constant. It is then worth
recalling again that this solution has been shown to be also an
attractor \cite{bonometto2012}.

Let us now consider the effects of Higgs' mass acquisitions. In fact,
at the Higgs' scale, the Lagrangian (\ref{interaction}) shall become:
\begin{equation} 
\label{HiggsLagrangian}
\tilde {\cal L}_m = -[\mu f(\Phi /m) + \tilde \mu ] \bar \psi \psi
\equiv -\mu [\exp(-C\Phi) + \tilde \mu/\mu] \bar \psi \psi~,
\end{equation}
so violating the CI invariance requirements. {Accordingly, below
 the Higgs' scale, the $\Phi$ Lagrangian is made by the free
 Lagrangian (\ref{freeLagrangian}) and the above term
 (\ref{HiggsLagrangian}).} As a matter of fact, such violation shall
become effective only when the $\tilde \mu$ matches the $\mu f$
term. However, by re-doing the functional differentiation in
Equation~(\ref{eulerolagrange}), we also obtain:
\begin{equation}
{\delta \tilde{\cal L}_m \over \delta \Phi} =-\frac{f'(C\Phi
 )}{f(C\Phi)+{\tilde \mu/\mu}}\rho_c = {C \over 1 + {\cal R}
 \exp[C(\Phi-\bar \Phi)]} \rho_c ~.
\label{variablecoupling}
\end{equation}

Here, $ {\cal R} = (\tilde \mu/\mu) \exp (C \bar\Phi)~, $ $\bar\Phi$
being the value of the field at a suitable reference time, e.g., during
the stationary regime. Let also $\exp(C\bar \Phi) = \mu/(g_h m_p)$.
We can then outline that, if the reference time is changed from $\bar
\tau_1$ to $\bar \tau_2$, both assumed to belong to the C.I. expansion
epoch, it shall be:

\begin{equation}
{\cal R}_1 = {\cal R}_2 {\bar \tau_1}/{\bar \tau_2}~~~~{\rm and}~~~~
g_{h,1} = g_{h,2} {\bar \tau_2}/{\bar \tau_1}~.
\end{equation}

The key point, however, is that the dynamical equations, even in the
presence of a Higgs' mass for the $\psi$ field, keep the form
(\ref{nonrelativisticeq1}), once we replace:
\begin{equation}
 C \to C_{eff} = {C \over 1 + {\cal R} \exp[C(\Phi-\bar \Phi)]}~~~{\rm
 and/or}~~~ \beta \to \beta_{eff} = {\beta \over 1 + 
 {\cal R} \exp[C(\Phi-\bar \Phi)]}~.
\label{newC}
\end{equation}

Then, when the $\Phi$ increase causes $\Phi-\bar \Phi$ to approach
$-\ln({\cal R})/C$, the denominators in Equation~(\ref{newC}) differ from
unity, so leading to a substantial violation of the primeval CI and
to suppressing the effective coupling intensity.

By assuming $g_h = 2\pi$ and a mass $\mu = 115\, $eV, we work out the
dependence of $\beta_{eff}$ on $a$ shown in Figure \ref{figure 2}, for the same
SCDEW model of Figure \ref{figure 1}.

Let us conclude this subsection by considering the $\Phi$ transition
from kinetic to potential. In low-$\beta$ coupled-DM approaches, the
transition occurs at a redshift dependent on some parameter(s) included
in the potential. In turn, one must tune such parameter(s), so as to
allow a fair amount of DE at $z=0$.

When directly dealing with the state parameter, we shall similarly
require:
\begin{equation}
w(a) = {1 - A \over 1 + A} ~~~{\rm with} ~~~ A = \left(a \over a_{kp}
\right)^\epsilon~,
\label{kp}
\end{equation}
so to obtain the kinetic-potential transition at a redshift $z_{kp}
\simeq a_0/a_{kp}-1$, then yielding a fair amount of DE at
$z=0$. Results are scarcely dependent on the exponent $\epsilon$,
whose arbitrariness somehow mimics the arbitrariness in the potential
choice.

\begin{figure}[H]
\vskip -1.8truecm
\centering
\includegraphics[width=6cm]{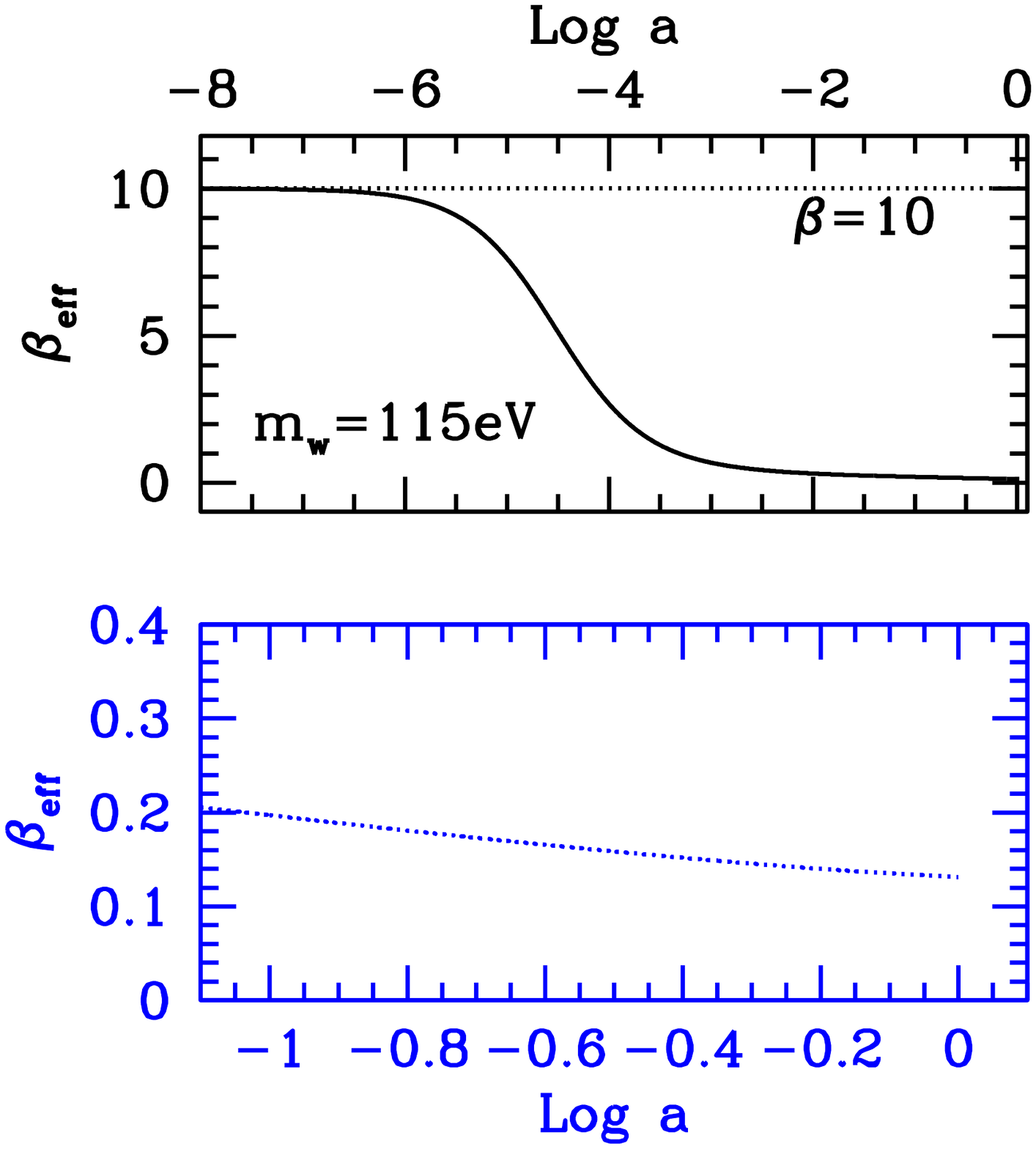}
\vskip -1.8truecm
\caption{As a consequence of Higgs' mass acquisition, the coupling
 constant decreases, about the redshift when warm DM turns
 non-relativistic. In the bottom blue plot, we show values close to
 $z=0$, ranging between 0.08 and 0.2. Such would be well within the
 observational range, even though all DM was coupled. }
\label{figure 2}
\end{figure} 

\subsection{Linear Fluctuation Evolution}
\label{lf}

Linear fluctuations in SCDEW models were first discussed in \cite{bonometto2014},
starting from initial conditions set out of horizon. In a synchronous
gauge, the metric shall then read:
\begin{equation}
 \label{newmetric}
ds^2 = a^2(\tau) [d\tau^2 - (\delta_{ij}+h_{ij}) dx_i dx_j)]~,
\end{equation}
$\tau$ being the conformal time, while gravity perturbation is
described by the three-tensor $h_{ij}$, whose trace is $h$. Let then:
\begin{equation}
\phi = \Phi + {b \over m_p} \varphi~,
\label{varphi}
\end{equation}
be the sum of the background field $\Phi$ considered in the previous
section and a perturbation described by $\varphi$. The most delicate
issue concerns the treatment of the $\varphi$ field, whose standard
equations include the second derivative of a $\Phi$ self-interaction
potential. If we wish to keep to the approach of setting $w(a)$,
instead of the potential, we must then replace:
\begin{equation}
2V'' = {A \over 1+A} \left\{ {\dot a \over a} {\epsilon \over 1+A}
\left[ \epsilon_6 {\dot a \over a^3} + 2C {\rho_c \over \dot \Phi}
\right] + \left[ {\dot a \over a^3} {\ddot \Phi \over \dot \Phi}
+ {d \over d\tau} \left( \dot a \over a^3 \right) \right]
\epsilon_6 + 2C {\dot \rho_c \over \dot \Phi} \right\}
\end{equation}
with $A$ and $\epsilon$ defined as in Equation~(\ref{kp}). Here, $\epsilon_6
= \epsilon-6$, while $\rho_c$ is the background density of coupled-DM.

It should be however outlined that most of the arguments of this work
concern the period of CI expansion, when these detailed behaviors are
marginally relevant. Let us however outline that, by using a linear
program we built, starting from this analysis, one can soon appreciate
that CMB anisotropies and polarization spectra, in SCDEW, exhibit just
tiny differences from LCDM, whose size further decreases when greater
$\beta$'s are considered. For instance, for $\beta \sim 10$,
discrepancies are safely below 1$\, \%$.

\section{A Top-Hat Fluctuation in the Early Universe}
\label{sec 3}

Let us then consider an overdensity, entering the horizon with an
amplitude $\delta_{c,hor}$, in the very early universe. We shall
mostly assume that $\delta_{c,hor} > 0$ and $\simeq 10^{-5}$,
i.e., that it is close to the top likelihood value for positive
fluctuations, with a Gaussian distribution.

The approach described below works only for $\delta_{c,hor}$ values
small enough to allow $\delta_c$ to enter a non-linear regime when
already non-relativistic. The ``rare'' case of $\delta_c$ entering
the non-linear regime when still relativistic would be relevant for
predictions on primeval black holes (see, e.g., \cite{carr2010} and the
references therein) and will be discussed elsewhere.

The critical point is illustrated in Figure \ref{figure 3}. The linear program
shows a progressive growth of the coupled-DM fluctuation. The growth
rate is greatest in the (linear) relativistic regime, at horizon
crossing, but is never discontinued, and in the non-relativistic
regime, $\delta_c \propto a^\alpha$ with $\alpha \simeq 1.6\, .$ The
process occurs in spite of DM being $\sim$1\% of the total
density, while still in the radiation-dominated~epoch.

\begin{figure}[H]
\centering
\includegraphics[width=6.5cm]{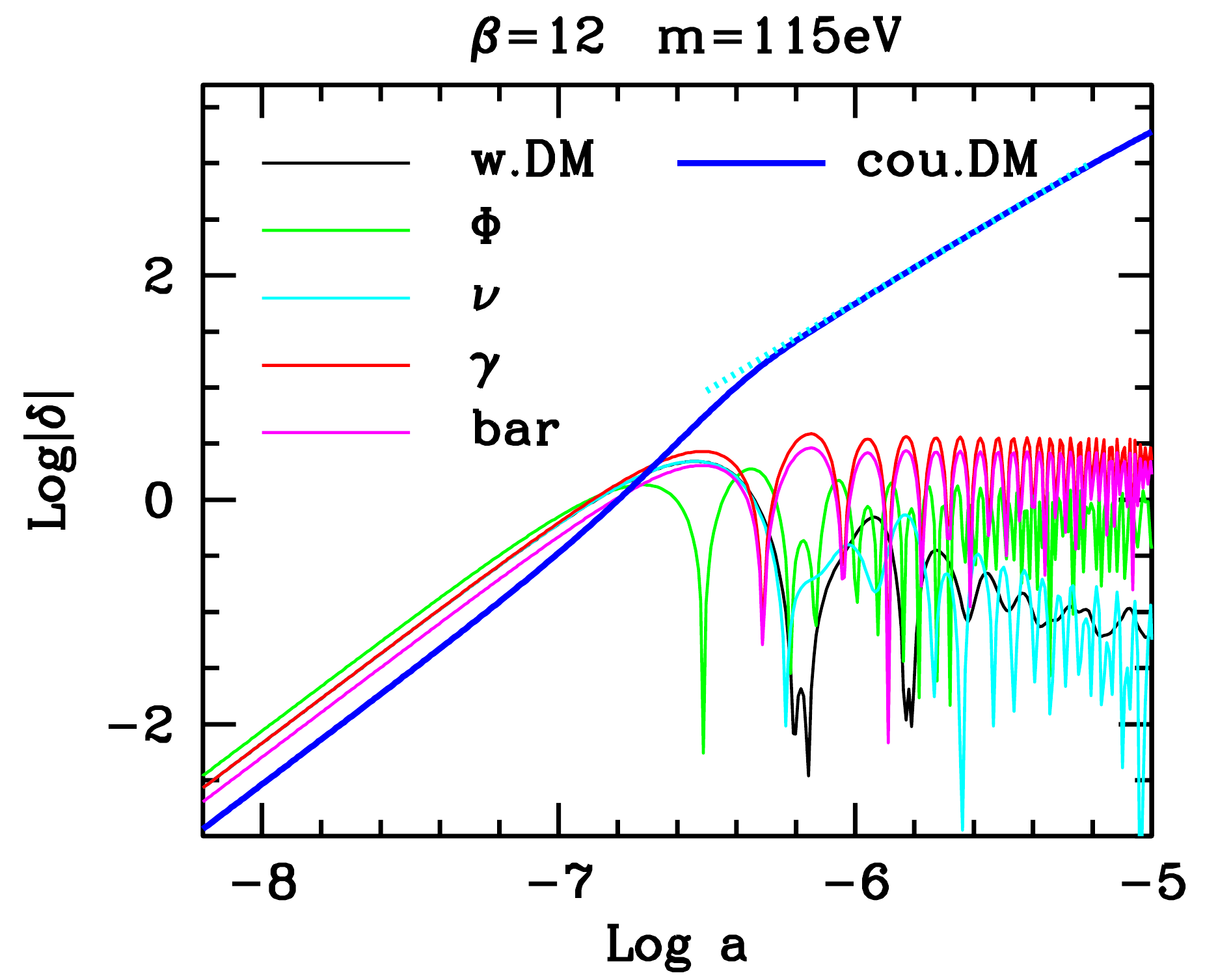}
\caption{Fluctuation evolution in the cosmic components at their entry
 in the horizon. For numerical reasons, we evaluated it at the end of
 the effective CI expansion, but the shape of the plot is invariant
 for shifts along the $a$ axis. The dotted line corresponds to a
 steepness $\alpha=1.6$. }
 \label{figure 3}
\end{figure} 

{As we shall soon see,}
this behavior, however, can be straightforwardly understood, on the
basis of the Newtonian limit of coupled-DM dynamics, as discussed in
\cite{maccio2004} (see also \cite{baldi}), when first aiming to
perform $N$-body simulations of coupled-DE models.

{Let us however first outline why this behavior is important and
 what problems it leads to. This is illustrated in Figure \ref{glevo},
 where we extend the plot of fluctuation evolution down to $z=0$, for
 all cosmic components.

\begin{figure}[H]
\centering
\includegraphics[width=6.5cm]{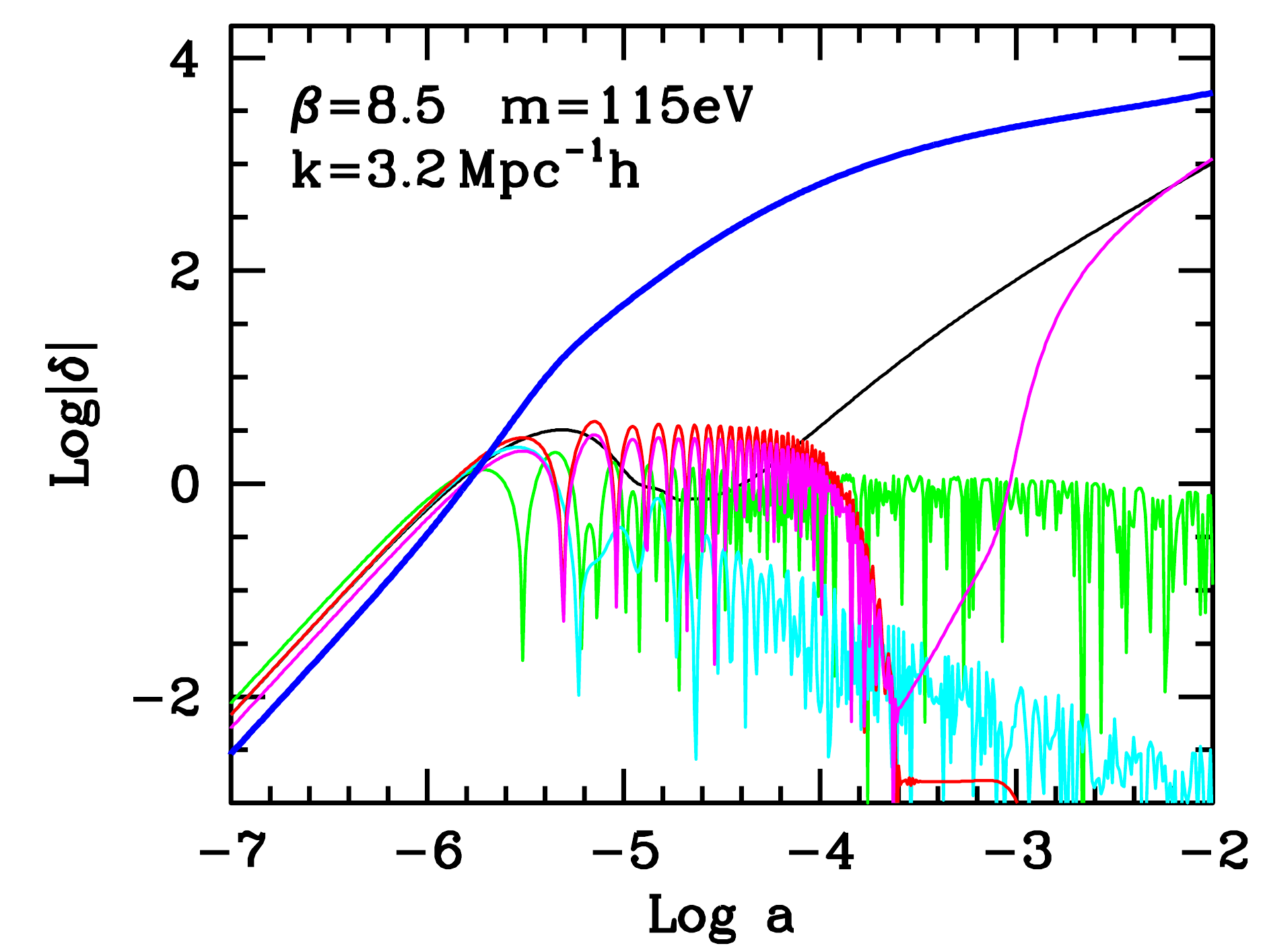}
\caption{Fluctuation evolution in the cosmic components from their
 entry in the horizon until $z = 0$, for the model and scale indicated in the
 frame. Model parameters are selected so as to cause an early coupled-DM
 non-linearity on such a scale. Colors as in the previous figure. }
\label{glevo}
\end{figure} 

 The role of coupled-DM fluctuations (in blue) is then evident: they
 revitalize warm DM and baryon fluctuations, as soon as warm DM
 de-relativizes and baryons decouple from radiation. In ordinary warm
 DM models, no structures are expected on any scale entering the
 horizon before warm DM de-relativizes. In SCDEW models, it is no
 longer, so: coupled-DM fluctuations survive until warm DM is able to
 accrete onto them.

 Henceforth, the fact that coupled-DM fluctuation undergoes an early
 growth is vital to allow SCDEW models to yield results close to
 LCDM.

 Figure \ref{glevo} allows us to see also which problems may arise.
 Being based on a linear algorithm, results do not depend on
 fluctuation normalization; $k=3.2\, h\, $Mpc$^{-1}$, however,
 corresponds to a big galaxy mass scale, being expected, on average,
 to turn non-linear at $z \sim 0$. Figure \ref{glevo} then shows
 that coupled-DM fluctuations have been non-linear since $z \sim
 10^{3}$--$10^4$. Accordingly, predictions based on a purely linear
 algorithm are just approximate as, surely, $\delta \rho_c$ cannot
 exceed $\rho_c$.

 A model with $m_w =115\, $eV and $\beta \sim 8$--9 was selected for
 this figure, so as to emphasize the problem, which for such $m_w$,
 occurs just on smaller mass scales if greater $\beta$ values
 (typically $>10$) are preferred. The problem however exists, and the
 need to explore the effects of coupled-DM fluctuations, when turning
 non-linear, needs to be approached. This makes the analysis of
 spherical overdensities so significant.

 Let us however first outline how \cite{maccio2004} understand the
 coupled-DM fluctuation behavior.}
They showed that
coupling effects are equivalent to: (i) An increase of the effective
gravitational push acting between DM particles, for the density
fraction exceeding average (while any other gravitational action
remains normal). The increased gravitation occurs as though $G =
1/m_p^2$ becomes:
\begin{equation}
G^* = \gamma\, G ~~~~{\rm with} ~~~~ \gamma = 1 + 4\beta^2/3
\label{gstar}
\end{equation}
(ii) As already outlined in Equations~(\ref{rhoc})
and (\ref{integre}), coupled-DM particle masses progressively
decline. This occurs while the second principle of dynamics still
requires that ${\bf f} = {\bf p}'$ (here, the prime indicates
differentiation with respect to the ordinary time $t$). This yields the
dynamical equation:
\begin{equation}
{d{\bf v} \over d\, t} = {{\bf f} \over m_{eff}} + \left|m_{eff}' \over
m_{eff} \right| {\bf v}~,
\label{fuma}
\end{equation}
i.e., an {extra push} to particle velocities, adding to the
external force~{\bf f}. Once Equations~(\ref{gstar}) and (\ref{fuma}) are
applied, the all of effects of coupling are taken into account; in
particular, the (small) $\Phi$ field perturbations cause no effect
appreciable at the Newtonian level (see again \cite{maccio2004,baldi}).

The self-gravitational push due to $\delta_c$ is then proportional to:
\begin{equation}
 G^* \delta_c \rho_c = G\, \delta_c \rho_{cr} {1 \over 2 \beta^2} \times
 \left(1 + {4\beta^2 \over 3} \right) =
 G\, \delta_c \rho_{cr} 
 \left( {2 \over 3} + {1 \over 2\beta^2} \right),
\end{equation}
as though concerning the whole critical density $\rho_{cr}$ at that
time, although with an amplitude reduced by a factor (slightly
exceeding) 2/3. However, also the impact of this factor $\cal O$ (unity) is
secondary, with respect to the effects of the {extra push} due to
particle mass decline. 

Such a fast increase will eventually lead $\delta_c$ into the
non-linear regime. In order to explore the dynamics in such a regime,
we can assume that the fluctuation is a spherical top-hat density
enhancement, of amplitude $\delta_c$. This is clearly an {ad hoc}
assumption, as real top-hat fluctuations can be expected to be
extremely unlikely. It is however a pattern widely followed in the
literature to explore the non-linear behavior, by passing from a
Lagrangian to a Eulerian perspective and by using then the equations
obeyed by the top-hat radius $R = ca$ ($c:$ comoving top-hat radius),
which in general, allows an easy integration. In suitable models and
epochs, there are even cases when an analytical integration is
possible \cite{press}.

The relation between $c = R/a$ and the density contrast $\Delta_c =1 +
\delta_c$ then reads:
\begin{equation}
\Delta_c = 1+\delta_c = \Delta_{c,r} c_r^3/c^3~,
\end{equation}
as the subscript $_r$ refers to a suitable reference time;
accordingly, by assuming $\delta_c \propto \tau^\alpha$,
\begin{equation}
{\dot c \over c_r} = -{\alpha \over 3} {\delta_{c,r} \over \Delta_{c,r}}
{1 \over \tau}
\label{dotc}~;
\end{equation}
this relation allows us to chose arbitrarily the time $\bar \tau$ when
we start to use $c$ instead of $\delta_c$ to follow the top-hat
dynamics, provided we do so in a fully-linear regime.

In the next section, we shall provide the equations fulfilled by $c$
and integrate them, following the pattern illustrated by
\cite{bonometto2017b}. In turn, this pattern is based on previous
papers \cite{mainini2005} (see also \cite{mainini2006}), concerning
top-hat evolution in (weakly) coupled-DE models. This is however
necessary to enable us to upgrade the discussion on top-hat
virialization.

\section{Top-Hat Dynamics}
 \label{sec4}
 
 In this paper, we shall however restrict the analysis to the very
 early universe, during the C.I. expansion era. We shall later comment
 on the relevance of the results concerning this era and on their
 extension to later periods.

By following the arguments leading to Equation~(9) in \cite{mainini2005},
the evolution of the overdensity can be shown to follow the equation:
\begin{equation}
\ddot c = -( {\dot a / a} - C\dot \Phi ) \dot c -\gamma
G [M(<R) - \langle M(<R) \rangle]/( ac^2)~.
\label{eq9}
\end{equation}

Differentiation is with respect to $\tau$; $\Phi$ is the background
scalar field; $M(<R)$ is the mass within $R$, while $\langle M(<R)
\rangle$ is the {average} mass in a sphere of radius $R$.
According to Figure \ref{figure 3}, we can reasonably assume all components, but
coupled-DM, to be unperturbed. Then, while:
\begin{equation}
G \langle M_c(<R) \rangle = {4\pi \over 3} G\rho_{cr} \Omega_c a^3 c^3
= {\Omega_c \over 2 \tau^2} a\, c^3~,
\label{GbarM}
\end{equation}
as, during the C.I. expansion era,
\begin{equation}
{8 \pi \over 3} G \rho_{cr} a^2 = {1 \over \tau^2} ~.
\label{h2}
\end{equation}

Accordingly:
\begin{equation}
{1 \over \bar \Delta} G M_c(<R) = G {4\pi \over 3} m_{eff}(\tau)\,
\bar n_c\, \bar a^3 \bar c^3 = G {4\pi \over 3} m_{eff}\, n_c ~a^3
\bar c^3 = {\Omega_c \over 2 \tau^2} a\, \bar c^3
\label{1od}
\end{equation}
(all ``barred'' quantities refer to the ``initial'' time $\bar \tau$);
$n_c$ is the number density of coupled-DM particles; although their
mass $m_{eff} \propto \tau^{-1}$, and the comoving number $n_c a^3$ is
constant in time. Therefore, 
\begin{equation}
 {G \over ac^2} [M(<R) - \langle M(<R) \rangle] = {\Omega_c \over 2
 \tau^2 x^2} \bar c \left({\bar \Delta}-x^3 \right)~.
\label{2term}
\end{equation}

In turn, the difference $h_0 = \dot a/a-C\dot\Phi$ exactly vanishes,
during the early CI expansion, both terms being then $ 1/\tau,$ so that:
\begin{equation}
x'' = - g \left(\bar \Delta - x^3 \right) {1 \over u^2 x^2}
\label{new9}
\end{equation}
with {$''$ indicating double differentiation with respect to $u =
 \tau /\bar \tau$}, and:
\begin{equation}
g = \gamma \Omega_c/2 = 1/3+1/(4\beta^2)
\label{g}
\end{equation}
increasingly close to 1/3, as $\beta$ increases. Let us outline that
this equation is visibly self-similar during the CI expansion,
including just quantities expressing a ratio between actual and
initial values.

\section{Virialization}
 \label{sec5}
 
Numerical solutions of Equation~(\ref{new9}) yield the expected growth and
successive re-contraction of the top-hat radius $R$, while the density
contrast $\Delta_c$ gradually increases. An {ideal} top-hat would
expand and re-contract down to a relativistic regime, and this is
obtainable from the Equation~(\ref{new9}) integral.

Minimal deviations from sphericity, however, which do not matter
during expansion, become determinant when $R$ starts to decrease. The
potential energy $U_c(R)$ is however set by the radius $R$, and also
kinetic energy $T_c(R)$ is hardly modified, although the direction of
motions gradually loses coherence, as they are no longer just directed
toward the center of the sphere.

Accordingly, the expression:
\begin{equation}
T_c(R) = {3 \over 10} M_c \left( dR \over dt \right)^{\, \, 2}~,
\label{Tc}
\end{equation}
keeps invariable (the factor 3/10 derives from integration on a
supposedly homogeneous sphere) while passing to differentiation with
respect to conformal time and to comoving variables,
\begin{equation}
 { dR \over dt} = {1 \over a}{ dR \over d\tau} = \dot c + {\dot a
 \over a} \, c\, ,
\end{equation}
so that
\begin{equation}
 2 \times {5 \over 3} {T_c \over M_c} \times { \bar \tau^2 \over \bar
 c^2 } = \left(x' + {1 \over u} x \right)^2~.
\label{kinetic}
\end{equation}

The potential energy is then made of two terms, arising from DM
fluctuation interacting with DM background and all backgrounds
interacting with themselves. Therefore, in agreement with
\cite{mainini2005,mainini2006}, we obtain:
\begin{equation}
{U(R) \over M_c} = -{3 \over 5} G {[\langle M_c \rangle + \gamma
 \Delta M_c] \over R } -{4 \pi \over 5}G \rho_{back} R^2 = -{3
 \over 5} \gamma G { \Delta M_c \over R } -{4 \pi \over 5}G
 \rho_{cr} R^2~.
\label{Ur}
\end{equation}
and from here, proceeding as we did for Equations~(\ref{GbarM}) and
(\ref{2term}), we finally obtain:
\begin{equation}
 {5 \over 3} {U_c(R) \over M_c} \times {\tau^2 \over \bar c^2} = - \left[{
  g x^2 \over 2} + {1 \over x} (\bar \Delta -x^3 ) \right].
\label{potentia}
\end{equation}

The virialization condition $ 2 {T_c } + {U_c } = 0 $ leads then to
requiring that:
\begin{equation}
(u x' + {g^{1/2}} x)^2 - (\bar \Delta/x
-x^2 ) -g x^2/2 = 0~.
\label{virial1}
\end{equation}

From the $c_v$ and $\tau_v$ values fulfilling this equation, we then
derive the {virial radius} $R_v = c_v a_v$.

The procedure described up to now is based on the results
of \cite{bonometto2017b}.
When comparing it with that paper, however, be aware of the different
meaning of the differentiation indicated by $'$ and $''$, which in~\cite{bonometto2017b}, were performed with respect to ordinary time.

\section{A Numerical Example}
 \label{sec6}
\begin{figure}[H]
\centering
\includegraphics[width=6.5cm]{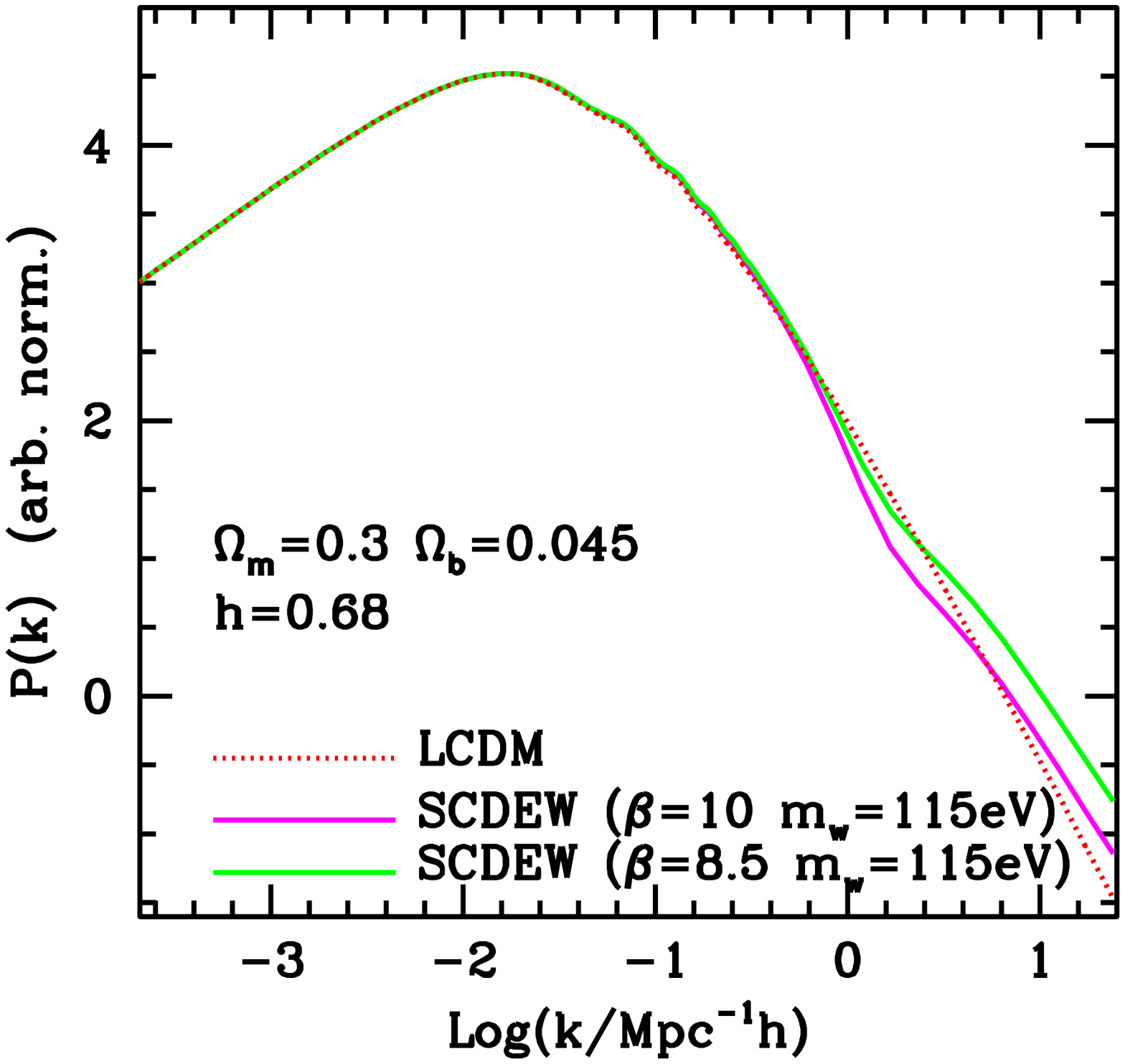}
\caption{Two SCDEW model spectra are compared with LCDM, at $z=0$. At
 high $k$, SCDEW cosmologies exhibit a higher spectrum, so indicating
 an earlier formation of low mass-scale structures. The ``green''
 model is then selected to provide an example of high-$z$ top-hat
 fluctuation evolution.}
\label{Pk8e10}
\end{figure} 
 
In Figure \ref{Pk8e10}, we compare the LCDM fluctuation spectrum with those of
the SCDEW model of Figure \ref{figure 1} and a further model with $\beta=8.5$; for
all models, the primeval spectral index $n_s=0.96$.

The latter SCDEW model is then selected to test the evolution of
spherical density enhancements, according to the equations discussed
in the previous section.

In Figure \ref{RDvst85}, we therefore plot the time dependence of the density
contrast $\Delta$ and the top-hat fluctuation radius $R$. In the
latter case, we also show its theoretically-expected behavior down to
full re-collapse and re-entering a relativistic regime: $R$ and
$\Delta$ are divided by their ``initial'' values $R_{in}$ and
$\Delta_{in}=1+10^{-3}$; during the C.I. expansion era, the expected
behaviors are however independent from the initial redshift $z_{in}$,
yielding the initial time $t_{in}$. These plots however hold for a
fluctuation entering the horizon with an amplitude $\delta_h \sim
10^{-5}$ (or smaller). We however warn the reader that non-average
fluctuations entering the horizon with a greater $\delta_h$ could
still be in the relativistic regime when their amplitude is $10^{-3}$.

\begin{figure}[H]
\centering
\includegraphics[width=6.5cm]{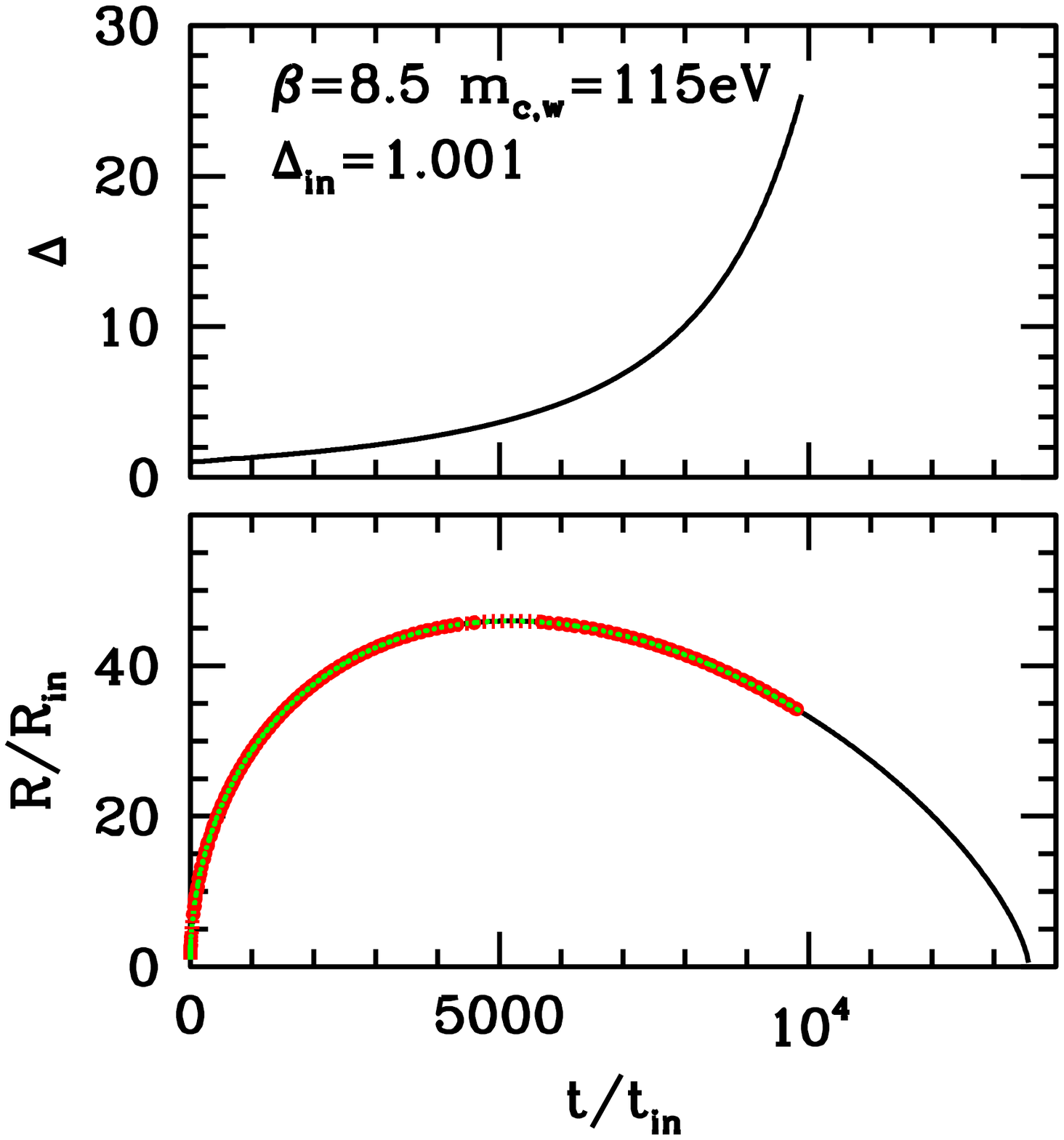}
\caption{Density contrast (top frame) and radius (bottom frame) of a
 top-hat fluctuation expanding and then re-contracting, in the
 non-relativistic regime. The density contrast is shown until the
 virialization condition is fulfilled, while the evolution of $R$ is
 plotted down to full re-contraction and entering a relativistic
 regime. Both $R$ and $\Delta$ are divided by their ``initial''
 values $R_{in}$ and $\Delta_{in}=1+10^{-3}$, so that the plots hold
 for any initial time $t_{in}$ during the C.I. expansion era.}
\label{RDvst85}
\end{figure} 

The results shown here are for a model different from the ones
discussed in \cite{bonometto2017a}. Moreover, here, we meant to
restrict ourselves to the dynamics occurring in the CI expansion era.
It is then quite significant that the difference between the actual
values of the virial density contrast $\Delta_v$ and the ratio
$R_v/R_{in}$, estimated here, exhibit a discrepancy <1\% from
those previously worked out. A similar conclusion concerns also
the times needed to reach virial equilibrium. The precise values
obtained here are reported in Table \ref{table 1}.

\begin{table}[H]
\caption{{ Virial time, radius and density contrast for different $\delta_{in}$.}} 
\centering
\begin{tabular}{ccccc}
\toprule
\boldmath{$\delta_{in}$}	& \boldmath{$\tau_v/\tau_{in}$}	& \boldmath{$t_v/t_{in}$} & \boldmath{$R_v/R_{in}$} & \boldmath{ $\Delta_v$}\\
\midrule
$10^{-3}$ & 99.4 & 9880 & 33.8 & 25.41 \\
$10^{-1}$ & 6.24 & 38.94 & 2.19 & 25.38 \\
\bottomrule
\end{tabular}
\label{table 1}
\end{table}

{In Table \ref{table 1}, besides of the values obtained if starting to follow the
spherical top-hat when $\delta=10^{-3}$, we report the values
obtainable if starting when $\delta=10^{-1}$. As previously outlined,
in fact, our treatment is slightly improved with respect to the one
proposed in \cite{press} and usually reported in textbooks, when it
was supposed that, initially, the overdensity expands coherently with
the Hubble flow. In fact, here, we work out the initial $\dot R$ from
the linear value of $\dot \delta$. For $\delta=10^{-3}$, this is just a
minimal correction. This however allows us to compare results obtained
by starting at different $\delta$'s, so estimating the non-linear
effects between such $\delta$ values.

Table \ref{table 1}, therefore, allows us to appreciate that non-linear effects
between $\delta=10^{-3}$ and $\delta=10^{-1}$ increase the virial density
contrast just by $\sim$0.12$\%$. This point is to be born in mind
for the later discussion on coupled-DM heating effects.}

\section{After Virialization}
 \label{sec7}
 
In previous sections, as well as in Figure \ref{RDvst85}, however, the approach to
virialization is treated in a schematic and idealized way. As a matter
of fact, to settle in virial equilibrium, the top-hat needs that
(tiny) deviations from full homogeneity existed. They hardly matter
until self-gravity just slows the overdensity expansion, but get
critical when the $\Delta$ increase causes $R$ to decrease.

Inhomogeneities cause then growing deviations from radial directions
of individual point velocities. They bear an effect also before the
virial density contrast $\Delta_v$ is attained, so that the time to
reach the virial radius $R_v$ exceeds the one computed above.
Randomized velocities are however still subdominant when $R_v$ is
attained, so that contraction is not discontinued, stopping only when
particle velocities become fully disordered.

Once this occurs, the system owns an excess kinetic energy with respect
to potential depth where it collapsed; therefore, $R$ starts to
re-increase towards a virial settling. In turn, this apparently
implies a partial reordering (with opposite sign) of velocities in the
radial directions. Accordingly, when the system re-approaches a virial
equilibrium condition from below, it will eventually bounce above
it. Altogether, a series of oscillations around the virial condition
is needed, before a possible system settling on it.

All of that is not different from what is expected to occur, after
recombination, for the evolution of a top-hat fluctuation in baryons
and dark matter (see, e.g., \cite{press}).

There is however a critical difference between such a case and the
present context. This point was partially outlined by
\cite{bonometto2017b}, but here, we shall show it in a more
quantitative fashion.

Let us then outline that, according to Equations~(\ref{Tc}) and (\ref{Ur}),
the top-hat virial $V$ is obtainable from the relation:
\begin{equation}
{5 \over 3} {V \over M_c} = \langle v^2 \rangle - {4\pi \over 3} G
\rho_{cr} \left( \gamma \Omega_c \Delta - {1 \over 3} + \Omega_c
\right)R^2~,
\label{wireq}
\end{equation}
once we replace ${R'}^2 = \langle v^2 \rangle$, i.e., once turning
coherent contraction into randomly-distributed speeds.

It is then convenient to multiply this relation by $m_{eff}^2$ and
outline the vanishing of the virial through the approximate relations:
\begin{equation}
\langle p^2 \rangle = \gamma G {N_c m_{eff}^3 \over R_v} =
{4\pi \over 3} \gamma G \rho_{cr,v} \Omega_c
\Delta_v R_v^2 m_{eff}^2 = g \left( R_v \over 2t_v \right)^2 \Delta_v 
m_{eff}^2 ~,
\label{wireq1}
\end{equation}
($N_c$ is the total number of coupled-DM particles, yielding a total
mass $N_cm_{eff}$). The reach of virial equilibrium should then imply
that the average momentum:
\begin{equation}
p_v^2 = \gamma G N_c m_{eff}^3 (\tau_v)/ R_v~,
\label{pv2}
\end{equation}
is conserved at any time $\tau > \tau_v$, when virialization occurred.

A similar equation can be written also for the case \cite{press}, the
main difference being that $m_{eff}$ then exhibits no time dependence.
Accordingly, in a such case, the system oscillations around virial
equilibrium occur while the average particle momentum $p_v$ also
maintains the momentum yielding virial equilibrium.

On the contrary, in our case, $p_v$ soon becomes a non-equilibrium
momentum as particles with kinetic energy $p_v^2/2m_{eff}$ are able to
evaporate from the fluctuation. A first estimate of the momentum kept
by evaporating particles is then provided by the last term in
Equation~(\ref{wireq1}), by using the numerical values in Table \ref{table 1}.
Accordingly:
\begin{equation}
 \label{pv2}
 {\langle p^2 \rangle \over ~ m_{eff}^2} \simeq
 3\, g \times 10^{-4} \left( R_{in} \over 2t_{in} \right)^2
\simeq 3\, g \times 10^{-4} {\tau_h \over \tau_{in}}~,
\end{equation}
by assuming $R_h = 2\, t_h$ (the suffix $_h$ refers to horizon
crossing). Notice also that $3g$ just slightly exceeds unity (see
{Equation}~{(\ref{g})}). 

The ratio $\tau_h/\tau_{in}\, $, evaluated by assuming $\delta \propto
\tau^{1.6}$ (as well as $|\delta_h| = 10^{-5}$), is $\sim$5.6$ \times
10^{-2}$. Accordingly:
\begin{equation}
 {\langle p^2 \rangle \over ~ m_{eff}^2} \sim 6 \times 10^{-6}
\label{pom}
\end{equation}
and this makes us sure that we are quite far from the relativistic
regime.

However, in order to come out from the virial ``potential well'',
particles necessarily loose momentum. In order to evaluate such
momentum reduction, we need to estimate the time a particle needs, on
average, to flow out from the well, as well as its depth at that time.

\section{Evaporation}
 \label{sec8}
 
It can be however shown that potential well evaporation is a rather
simple process, and its duration does not exceed the time taken by
a particle to cross the ``virialized'' lump much. In this section, we
explain why this is so.

Let then $p^2_v = \langle p^2 \rangle$, and let us notice that,
according to Equation~(\ref{wireq}) and assuming that the lump keeps
a size close to $R_v$, the {escape momentum} from the spherical
enhancement is:
\begin{equation}
 p_{esc}(t) = p_v (t_v/t)^{3/4}~,
 \label{pesc}
\end{equation}
so that, if all particles had exactly the average momentum,
evaporation would occur soon. In general, we can assume that
momenta are suitably distributed, so that only a fraction of particle
momenta exceed $p_{esc}$ at any time $t > t_v$. Their eventual
evaporation bears a dual effect: it lowers $p_v = (\langle p^2
\rangle)^{1/2}$, however causing also a loss of binding energy,
because of the mass loss. Of course, while the fastest particles
escape, the particle distribution tends to reassume its initial shape
$f(p/p_{top})$ ($p_{top}$: momentum corresponding to the top
distribution value), but let us keep to a schematically discrete
process and seek which fraction of high-speed particles may
conveniently escape, reducing $ \langle p^2 \rangle$ more rapidly than
the potential term $\gamma G N_c m_{eff}^2/R_v$.

If one assumes that the distribution has a Maxwellian shape:
\begin{equation}
 f(x) = 4\pi^{-1/2} x^2 e^{-x^2}~,
 \label{max}
\end{equation}
it is $p_{top} = (2/3)^{1/2} p_v$. Then, when all particles with $p >
\alpha p_{top}$ have evaporated, the average momentum reduces to:
\begin{equation}
 \langle p^2_\alpha \rangle = {2 \over 3}\, p_v^2 \int_0^\alpha dx\, x^2 f(x)
 \bigg/ \int_0^\alpha dx\, f(x)
\end{equation}
while the potential term reduces to:
\begin{equation}
 \gamma G {m_{eff}^2 \over R_v} N_c(\alpha) = p_v^2 {N_c(\alpha) \over N_c}
 = p_v^2 \int_0^\alpha dx\, f(x)~,
\end{equation}
$N_c(\alpha)$ being the ``residual'' particle number. The value of
$\alpha$ maximizing the ratio:
\begin{equation}
 F(\alpha) = {2 \over 3} \int_0^\alpha dx\, x^2 f(x)
 \bigg/ \left[ \int_0^\alpha dx\, f(x) \right]^2
\end{equation}
allows then the top advantage from progressive evaporation. For the
Maxwellian distribution (\ref{max}), such a minimum occurs for $\alpha_m
\simeq 1.21$ and $F(\alpha_m) \simeq 0.81\, .$

Equation~(\ref{pesc}) then tells us that $p_{esc}^2$ needs a time $t_{esc}
\sim 0.15\, t_v$ to decrease by this very amount. The questions are
then: (i) are the fastest particles able to stream out from the
overdensity within such time? (ii) are the residual particles able to
recover a Boltzmann distribution in that very time? These are
necessary conditions for the depleted system being able to recover a
virial equilibrium.

Both the streaming out and the reset times are safely greater than the
crossing time $t_{cross}$ a particle needs to cover a distance $R_v$
at a velocity $p_v/m_{eff}$. The last relation (\ref{wireq1}), also
reading $ p_v = (g\Delta_v)^{1/2} R_v m_{eff}/2t_v $, allows us to
compute:
\begin{equation}
 t_{cross} = 2t_v (g\Delta_v)^{-1/2} \sim 0.7\, t_v~,
 \label{tcross}
\end{equation}
according to the figures in Table \ref{table 1} and Equation~(\ref{g}). As $t_{cross}
\sim 5\, t_{esc}$, particles have no time to stream out from the
overdensity, let alone to rearrange into a renewed Maxwellian
distribution. In other terms: the escape momentum decreases too
rapidly. Henceforth, {no trapping effect is possible; particles
 simply flow out from the overdensity within a time $\sim$0.7$\,
 t_v$.}

It is then critical to determine which fraction of the momentum
(\ref{pom}) they shall be able to maintain. An estimate can be
performed by evaluating the difference $p_v-p_{esc}(t_{cross})$,
yielding an average reduction by a factor $\sim$1/3. The main
residual effect of the passage through an overdensity, then, is a sort
of {heating up} of the coupled-DM particles. As the horizon has
grown much greater than $R_v$, this {heat} is rapidly shared with
particles that did not belong to overdensities, being so reduced by a
further factor $\sim$1/2.

{Altogether, this leads us to estimate that evaporated coupled-DM
particles own a momentum:
\begin{equation}
 {\langle p^2 \rangle \over ~ m_{eff}^2 } \sim 10^{-6}~,
\label{pom1}
\end{equation}
with a reduction by a factor 1/6 with respect to (\ref{pom}). The
distribution of momenta, $g(p)$, may then be significantly different
from Maxwellian; however, the pressure and energy density of coupled-DM,
processed by overdensities, shall read:
\begin{equation}
 3\, P = \int d^3p\, \left(p \over p_0 \right)^2 p_0\, g(p)~,~~~
 \rho = \int d^3p\, p_0\, g(p)~,~~~
 \label{Prho}
\end{equation}
with $p^2_o = m_{eff}^2(1+p^2/m_{eff}^2)$, so that:
\begin{equation}
 3\, P = { \langle \langle p^2 \rangle \rangle \over ~ m_{eff}^2}
 \int d^3p\, p_0\, g(p)~~~~{\rm and} ~~~~~~~ w_c \sim {1 \over 3}\,
 10^{-6}
\end{equation}
is a sound estimate of the order of magnitude
 of its state parameter (in spite of the
average $\langle \langle ... \rangle \rangle$ being different from
$\langle ... \rangle$~). In linear analysis, such $w_c$ can be set to
zero, so making an error similar to numerically working in simple
precision.

Such velocity fields can cause some more consequences in the analysis
of top-hat density enhancements. In order to define the top-hat
itself, it should be:
\begin{equation}
 R_{in} {(da/dt) \over a} - {dR \over dt} \gg v_{th}~,
 \label{vth}
\end{equation}
i.e., the difference between the velocity due to the Hubble flow and
the top-hat growth velocity should substantially exceed the {thermal} velocities. Then:
\begin{equation}
 dR/dt = (x'+x/u)R_{in}/(2\, t_{in})
\end{equation}
and taking $v_{th}^2$ to be a fraction of the second term in
Equation~(\ref{pv2}), we obtain the condition:
\begin{equation}
 1-(x'+x/u) > 10^{-2}~,
 \label{meno2}
\end{equation}
assuring us that the condition (\ref{vth}) is fulfilled by a factor
$\sim$6.

In Figure \ref{vel}, we then plot the first term of Equation~(\ref{meno2})
and compare it with the growth of $\delta = \Delta-1$.
\begin{figure}[H]
\centering
\includegraphics[width=6cm]{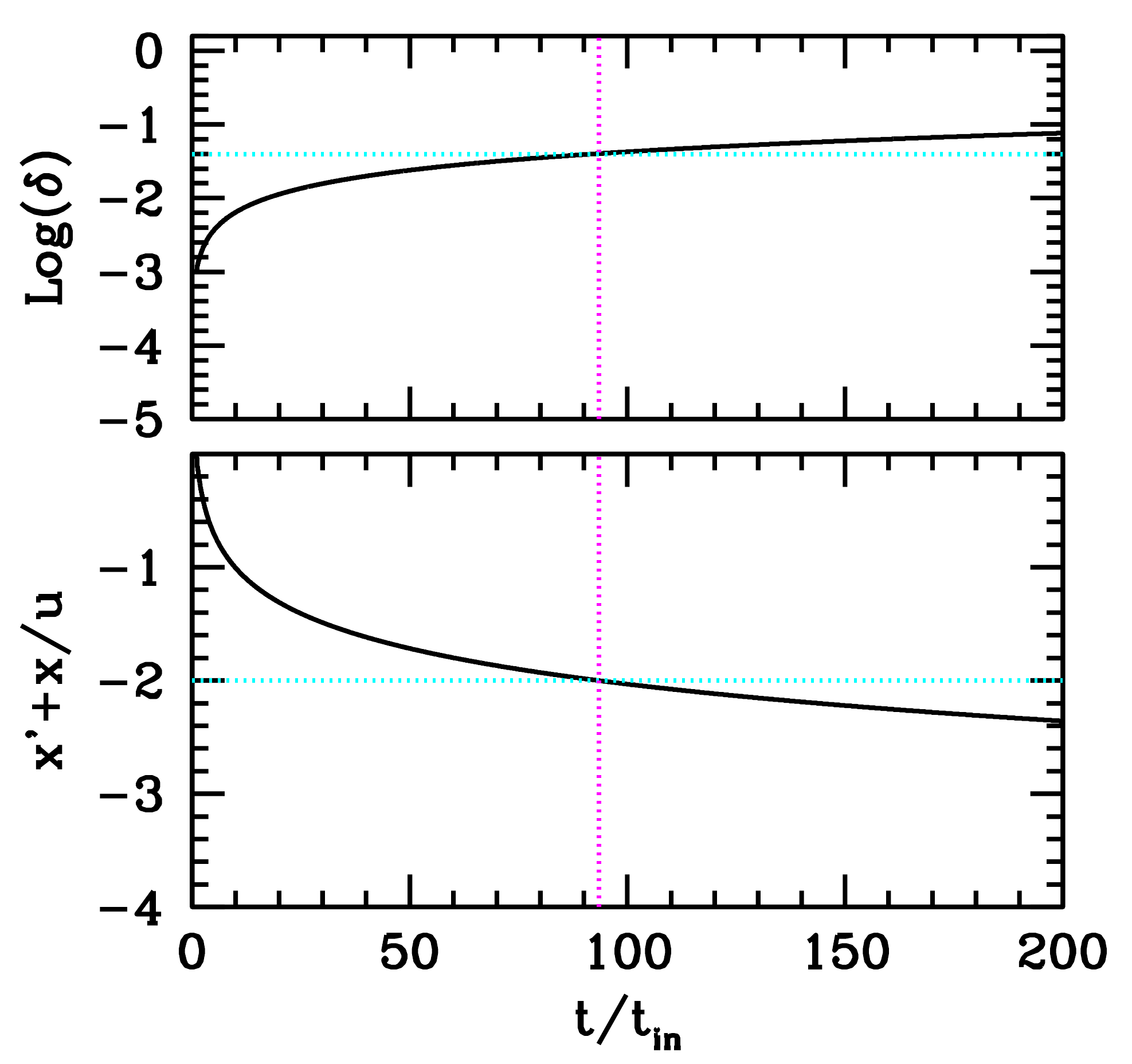}
\caption{The density contrast growth is compared with the size of
 coherent velocities at the boundary of the top-hat. In the figure,
 we single out the value of $\Delta$ when $|v|=v_{vir}$, before the
 top-hat reaches its full expansion.}
\label{vel}
\end{figure} 
It is then easy to estimate that, when the linear fluctuation $\delta
\simeq 0.1$, the boundary velocity of the top-hat exceeds ``thermal''
velocities by one o.o.m. or more.

This confirms that the essential difficulty to pass from a linear
Lagrangian description, based on density fluctuations $\delta$, to an
Eulerian description based on spherical top-hats is due to the
impact of ``thermal'' velocities in the very definition of top-hat
boundaries. Figure \ref{vel} however shows us that a top-hat can be
however safely defined when $\delta \sim 10^{-1}$, as then, the
top-hat growth yields velocities safely distant from the Hubble flow.

Passing from a Lagrangian to a Eulerian treatment when $\delta$ is so
``large'' causes the neglect of non-linear effects acting when
$\delta < 10^{-1}$. Table \ref{table 1} however showed us that the overall effect
disregarded is of the order of the permil. Accordingly, we are allowed
to conclude that a top-hat treatment keeps its substantial validity
also when coupled-DM has undergone a pre-heating process.}

\section{Discussion}
 \label{sec9}
 
This approximate estimate is based on assuming that density
enhancements (or depletions) are {spherical}. Realistic geometries
are surely quite different. The study of structure formation based on
top-hat fluctuations is known to allow for fair predictions.
Evaluating cosmic material ``heating'', however, is a different matter,
and it seems unclear whether we can claim a similar reliability.

Independent of our scheme, however, once particles are freed,
their momenta are subject to ordinary red shifting, with a rate however
equal to the $m_{eff}$ decrease. Henceforth, the ratio $p_{th}/m_{eff}$
keeps constant in time.

With the estimated level of ``thermal'' momentum $p_{th}$, however,
coupled-DM should be considered {cold}, as far as linear
fluctuation dynamics is concerned. This is so also because the
``pre-heating'' took place on a scale (significantly) smaller than
the size of fluctuations entering the horizon later on. Velocity
fields, absolutely non-relativistic, are unable to extend up to the
horizon scale. Over such larger scale, only the velocity fields
generated by the fluctuations shall matter.

{When we try to shift from the linear treatment to the top-hat
 dynamics, i.e., from a Lagrangian to a Eulerian scheme, $p_{th}$
 causes some problems as far as the very top-hat definition is
 concerned. We however showed that, if the shift occurs when $\delta
 \sim 10^{-1}$, coherent collapse velocities allow us to single out
 the top-hat with respect to the Hubble flow, in spite of ``thermal''
 motions. The {error} due to such a late Lagrangian--Eulerian
 transition has been also shown not to exceed the permil range. An
 effect of similar size could then be caused by ``thermal''
 velocities contrasting the gravitational push and slowing down the
 expansion stages, as well as the early re-contraction. On the
 contrary, the approach to virial equilibrium could be faster, as the
 mix up due to $p_{th}$ could facilitate the conversion of the
 coherent kinetic energy into disordered motions.}


Another question concerns the possibility of cumulative heating up,
when the same materials are involved in successive non-linearities,
over greater and greater scales, but indeed, the level of residual
``heat'' is however fixed by the capacity of materials to stream out
from the ``devirializing'' matter lump.

The whole question, however, appears rather intricate, and suitable
{ad hoc} simulations could be helpful to allow us a more complete
comprehension.

\section{Conclusions}
 \label{sec10}
 
In this paper, we aimed to continue the exploration of non-linear
effects in coupled-DM evolution, for SCDEW cosmologies. The basic
point is that coupled-DM fluctuations, however, grow when they enter
the horizon, in spite of the small density of coupled-DM, because of
the boost in its gravitational self-interaction.

This is the reason for the SCDEW model's success: coupled-DM fluctuations are
able to revive warm DM and baryon fluctuations, even on scales where
they were previously erased by free streaming or recombination,
allowing them to re-achieve an amplitude close to LCDM models.

This however could cause problems on small galactic and sub-galactic
scales, as coupled-DM fluctuations could enter an early non-linear
regime. How early this is depends on the model parameters. Let us outline
that this is essentially a technical difficulty, making it harder to
formulate exact model predictions, namely of the time when these
structure formed. This regime needs therefore to be explored, to go
beyond order of magnitude estimates, for these small scales.

Quite the same effect worked also earlier, over smaller and smaller
scales; in particular, even during the C.I. expansion era, coupled-DM
fluctuations exhibit a rapid growth and quickly enter a non-linear
regime.

This work is focused on this early effect, explored by considering
spherical top-hat overdensities. Previous papers
\cite{bonometto2017b} have shown that the density contrast they reach
at virialization is $\sim$ 25 times coupled-DM density, so that they
keep ``linear'' with respect to the overall cosmic density. Moreover,
it was shown that virialized structures are doomed to dissolve,
because the effective mass of coupled-DM particles undergoes a fast
decrease. In this paper, we however outline that, although freely
steaming from overdensities, coupled-DM particles are {heated up}
by the processing inside them.

This is the main new finding of this work. Heating causes particle
momenta to keep $\sim$15--20\% of their past and future {virial} value. In principle, this may interfere with the whole
treatment of overdensity evolution, which did not take into account
such a sort of {intrinsic} coupled-DM momenta. Quantitatively,
we however expect just a small effect.

Such an effect could be however important when we face the dynamics of
overdensities able to accrete warm DM particles and/or baryons. In
this case, we shall aim at precise numerical results, as they should
allow us to correct linear predictions on sub-galactic scales. This
analysis shall be performed in further work, although as already
outlined, safe quantitative results could only be derived through
suitable {ad hoc} simulations.

\vspace{6pt}

\authorcontributions{The two authors contributed to this paper in
 quite similar ways. Both authors have read and approved the final manuscript.}

\conflictsofinterest{The authors declare no conflict of interest.}


\begin{thebibliography}{999}
\bibitem[Riess(1998)]{riess}  
Riess, A.G.; Filippenko, A.V.; Challis, P.; Clocchiatti, A.; Diercks, A.; Garnavich, P.M.; Gilliland, R.L.; Hogan,~C.J.; Jha, S.; Kirshner, R.P.; et al. {
  Observational Evidence from Supernovae for an Accelerating Universe
  and a Cosmological Constant}. {\it Astron. J.} {\bf 1998}, {\em 116},
  1009--1038, doi:10.1086/300499.
  
\bibitem[Perlmutter(1999)]{perlmutter} 
Perlmutter, S.; Aldering, G.; Goldhaber, G.; Knop, R.A.; Nugent, P.; Castro, P.G.; Deustua, S.; Fabbro, S.; Goobar, A.; Groom, D.E.; et al. 
{Measurements of Omega and Lambda from 42 High-Redshift
  Supernovae}. {\it Astrophys. J.} {\bf 1999}, {\em 517}, 565--586, doi:10.1086/307221.
%

\bibitem[Kleidis(2015)]{kleidis2015}
 {Kleidis, K.; Spyrou, N.K.} 
 {Polytropic dark matter flows illuminate dark energy and  accelerated expansion}. {\it Astron. Astrophys.} {\bf
  2015}, {\em 576}, A23, doi:10.1051/0004-6361/201424402.
  
\bibitem[Kleidis(2015)]{kleidis2015a}
 Kleidis, K.; Spyrou, N.K. {Dark Energy: The Shadowy Reflection of Dark Matter? } {\it Entropy} {\bf
    2016}, {\em 18}, 94.
    
\bibitem[Kleidis(2015)]{kleidis2015b}
   Kleidis, K.; Spyrou, N.K.
     {Cosmological
    perturbations in the LCDM-like limit of a polytropic dark matter
    model}. {\it arXiv} {\bf 2016}, {arXiv:1707.08531}. 
%
\bibitem[Benitski(2017)]{Benisty2017} 
Benisty, D.; Guendelman, E.I. {Interacting Diffusive Unified Dark Energy and Dark
  Matter from Scalar Fields}. {\it Eur. Phys. J.} {\bf
  2017}, {\em C77}, 396, doi:10.1140/epjc/s10052-017-4939-x.
  
  \bibitem[Benitski(2017)]{Benisty2017a} 
Benisty, D.; Guendelman, E.I. {Radiation Like Scalar Field and Gauge Fields in
    Cosmology for a Theory with Dynamical Time}. {\it Mod. Phys. Lett. A} {\bf 2016}, {\it 31}, 1650188. 
%

\bibitem[Haba(2017)]{Haba2017} Haba, Z. {
  Thermodynamics of diffusive DM/DE systems}. {\it Gen. Relativ. Gravit.} {\bf 2017}, {\em 49}, 58, doi:10.1007/s10714-017-2224-9.
  
\bibitem[Bonometto(2012)]{bonometto2012} 
Bonometto, S.A.; Sassi, G.; La  Vacca, G. 
  {Dark energy from dark radiation in
    strongly coupled cosmologies with no fine tuning}. {\it J. Cosmol. Astropart. Phys.} {\bf 2012}, {\em
    08}, 015, doi:10.1088/1475-7516/2012/08/015.
    
\bibitem[Bonometto(2014)]{bonometto2014} Bonometto, S.A.; Mainini, R. {Fluctuations in strongly coupled cosmologies}. {\it  J. Cosmol. Astropart. Phys.} {\bf 2014},
  {\em 03}, 038, doi:10.1088/1475-7516/2014/03/038.
\bibitem[Bonometto(2015)]{bonometto2015} Bonometto, S.A.; Mainini, R.; Macci\'o,
  A.V. {Strongly coupled dark energy cosmologies: Preserving LCDM success and easing low scale problems---I. Linear theory revisited}. {\it Mon. Not. R. Astron. Soc.} {\bf 2015},  {\em 453}, 1002--1012, doi:10.1093/mnras/stv1621.
\bibitem[Maccio(2015)]{maccio2015} Macci\'o, A.V.; Mainini, R.; Penzo, C.;
  Bonometto, S.A. {Strongly coupled dark energy
    cosmologies: preserving LCDM success and easing low-scale problems---II. Cosmological simulations}. {\it Mon. Not. R. Astron. Soc.} {\bf 2015}, {\em 453}, 1371--1378, doi:10.1093/mnras/stv1680.
\bibitem[Bonometto(2017)a]{bonometto2017a} Bonometto, S.A.; Mezzetti, M.; Mainini,
  R. {Strongly Coupled Dark Energy with Warm dark
    matter vs. LCDM}. {\it arXiv} {\bf 2017}, {arXiv:1703.05139}.
    
\bibitem[Bonometto(2017)b]{bonometto2017b} Bonometto, S.A.; Mainini, R.
  {Growth and dissolution of spherical density enhancements in
    SCDEW cosmologies}. {\it J. Cosmol. Astropart. Phys.} {\bf 2017}, {\em 06}, 010, doi:10.1088/1475-7516/2017/06/010.
    
    
    
\bibitem[Maccio(2017)]{maccio2017} Macci\`o, A.V.; Frings, J.; Buck, T.;
  Penzo, C.; Dutton, A.A.; Blank, M.; Obreja, A. {The edge
    of galaxy formation I: Formation and evolution of MW-satellites
    analogues before accretion}. {\it arXiv} {\bf 2017}, arXiv:1707.01106.
%
\bibitem[Frings(2017)]{Frings2017} Frings, J.; Macci\`o, A.V.; Buck, T.;
  Penzo, C.; Dutton, A.A.; Obreja, A.; Blank, M. {The edge
    of galaxy formation II: Evolution of Milky Way satellite analogues
    after infall }. {\it arXiv} {\bf 2017}, arXiv:1707.01102.
%
\bibitem[Santos(2017)]{Santos2017} Santos-Santos, I.M.; Di Cintio, A.;
  Brook, C.B.; Macci\'o, A.V.; Dutton, A.; Domínguez-Tenreiro, R. {NIHAO XIV: Reproducing the observed diversity of dwarf
    galaxy rotation curve shapes in LCDM}. {\it arXiv} {\bf 2017}, arXiv:1706.04202.
%

\bibitem[Amendola(2010)]{amendola2010} Amendola, L.;  Tsujikawa, S. {\it Dark
  Energy}; Cambridge University Press: Cambridge, UK, {2010}; ISBN: 9780521516006.
  %
\bibitem[Bamba(2012)]{bamba} Bamba, K.; Capozziello, S.; Nojiri, S.; Odintsov, S.D. {Dark energy cosmology: the
    equivalent description via different theoretical models and
    cosmography tests}. {\it Astrophys. Space Sci.} {\bf 2012}, {\em 342}, 155--228, doi:10.1007/s10509-012-1181-8.


\bibitem[Amendola(2000)]{amendola2000} Amendola, L.
  {Coupled quintessence}. {\it Phys. Rev. D} {\bf 2000}, {\em 62}, 043511, doi:10.1103/PhysRevD.62.043511.


\bibitem[Carr(2010)]{carr2010} Carr, B.J.; Kohri, K.; Sendouda, Y.;  Yokoyama, J. { New cosmological constraints on primordial black holes}. {\it Phys. Rev. D} {\bf 2010}, {\em 81}, 104019, doi:10.1103/PhysRevD.81.104019.
\bibitem[Maccio(2004)]{maccio2004} Macci\'o, A.V.; Quercellini, C.; Mainini, R.; Amendola, L.; Bonometto,
S.A. {$N$-body simulations for coupled dark
  energy: Halo mass function and density profiles}. {\it Phys. Rev. D} {\bf 2004}, {\em 69}, 123516,
doi:10.1103/PhysRevD.69.123516.
\bibitem[Baldi(2010)]{baldi} Baldi, M.; Pettorino, V.; Robbers, G.; Springel, V.
{Hydrodynamical $N$-body simulations of coupled dark energy
  cosmologies}. {\it Mon. Not. R. Astron. Soc} {\bf 2010}, {\em 403}, 1684B, doi:10.1111/j.1365-2966.2009.15987.x.
\bibitem[Press(1974)]{press}  Press W.H.; Schechter, P. {Formation of Galaxies and Clusters of Galaxies by Self-Similar
  Gravitational Condensation}. {\it	Astrophys. J.} {\bf 1974}, 187, 425, doi:10.1086/152650.
\bibitem[Mainini(2005)]{mainini2005}  Mainini, R. {Dark matter-baryon segregation in the non-linear evolution of coupled dark energy model}.
{\it Phys. Rev. D} {\bf 2005}, 72, 083514, doi:10.1103/PhysRevD.72.083514.
\bibitem[Mainini(2006)]{mainini2006} Mainini, R.; Bonometto, S.A.
  {Mass functions in coupled dark energy models}. {\it Phys. Rev. D} {\bf 2006}, 74, 043505,
  doi:10.1103/PhysRevD.74.043504.

\end{thebibliography}
\end{document}